\documentclass[aps,preprintnumbers,eqsecnum,amsmath,amssymb,nofootinbib]{revtex4}  
\usepackage{graphicx}
\usepackage{bm}
\usepackage{epsfig}
\usepackage{color} 
\usepackage{float}

\begin{document}
\title{Triangle Feynman diagram in the timelike region}
\author{Mikhail A.~Ivanov$^a$, Dmitri Melikhov$^{a,b,c}$, and Silvano Simula$^d$}
\affiliation{
$^a$Joint Institute for Nuclear Research, Dubna, 141980, Russia\\
$^b$D.~V.~Skobeltsyn Institute of Nuclear Physics, M.~V.~Lomonosov Moscow State University, 119991, Moscow, Russia\\
$^c$Faculty of Physics, University of Vienna, Boltzmanngasse 5, A-1090 Vienna, Austria\\
$^d$INFN, Roma Tre, Via della Vasca Navale 84, I-00146, Rome, Italy
}
\date{\today}
\begin{abstract} 
In QFT, triangle Feynman diagram  $F(p_1^2,p_2^2,p_3^2|m_1,m_2,m_3)$ is an analytic function 
of its variables, whose analytic structure is fully determined by the location of singularities of the
propagators of particles in the loop.The form factor $F(p_1^2,p_2^2,p_3^2|m_1,m_2,m_3)$ is easily calculable in the
Euclidean region $p_i^2<0$, $i=1,2,3$. A rigorous way to obtain the form factor in the Minkowki region
is to perform the analytic continuation from the Euclidean region. We propose here a different approach.
There is a representation of the triangle as integral over Feynman parameters. In the Euclidean region  $p_i^2<0$, the integration over the Feynman parameters runs over segments of the real axis.
We show that the same feature holds also in the Minkowski region of the external momenta above the cuts:
The integration over the Feynman parameters still runs along the same segments of the real axis.
No deformation of the integration paths is forced by possible migrations of poles from the unphysical
sheet to the physical one. We demonstrate that all known rigorous results of dispersion representations
in the regions where some of the variables $p_i^2$ are in the physical Minkowski region above the cuts,
are reproduced by
the Feynman-parameter representation for $F(p_1^2,p_2^2,p_3^2|m_1,m_2,m_3)$ through the replacement 
$p_i^2 \to p_i^2 + i0$. Our observation opens the possibility to obtain the generic triangle diagram (e.g.
with fermions in the loop and with vertices containing derivatives) in the Minkowski region above the cuts
numerically by a simple evaluation of the Feynman-parameter integral over the segments of the real axis. 

\end{abstract}
\maketitle

\section{Introduction}
Triangle diagrams have many applications in quantum field theory:  
They give form factors of a particle like quark or electron;
they describe weak and electromagnetic form factors of hadrons and other bound states;
they provide essential contributions to amplitudes of hadronic decays; 
they are responsible for quantum anomalies.

Triangle Feynman diagram $F(p_1^2,p_2^2,p_3^2|m_1,m_2,m_3)$, notations as in Fig.~\ref{fig:0} 
(we also use the term form factor for this quantity)
is an analytic function of its variables, whose analytic structure is
fully determined by the location of singularities of the propagators of particles in the loop.
$F(p_1^2,p_2^2,p_3^2|m_1,m_2,m_3)$ is easily calculable in the Euclidean region of all variables, $p_i^2<0$.
A rigorous way to obtain the form factor $F(p_1^2,p_2^2,p_3^2|m_1,m_2,m_3)$ in the timelike region is to perform the
analytic continuation from
the Euclidean region in the variable(s) of interest, using dispersion representations
(see the original works \cite{nakanishi,karplus,landau,anisovich_1,fronsdal,norton,anisovich_2}
and the monograph \cite{burton}).
This is however a cumbersome procedure, which requires a detailed analysis of
migration of singularities (located on the unphysical sheet
for Euclidean external momenta) to the physical sheet as soon as some of the external momenta become positive.
This migration of singularities leads to the modification of the integration contours in the complex plane and
to the appearance of the anomalous thresholds and the anomalous cuts. An exhaustive analysis of the single and
the double dispersion representations of the triangle diagram for all
values of the external and the internal masses can be found in \cite{fronsdal}. These dispersion representations are however rather cumbersome and not easy to use for numerical estimates of the triangle diagram
in the Minkowski region \cite{lms,procura}. 
For extensive applications of dispersion representations in hadron physics we refer to 
\cite{k3pi,procura,oka,mh,as,mm,abreu2015,guo2,ims2020,guo2020,han2022,ms2021,kubis2024,gubernari2024,isidori2025,kubis2026} and references therein. 

An explicit expression for the scalar triangle diagram in terms of the dilogarithmic function 
was obtained in \cite{tHooft:1978jhc}. One can use this analytic expression to obtain the triangle in the Minkowski region.
Still, some caution is required in order to stay at the proper branches of the multi-valued dilogarithnic
function when moving to the Minkowski region. Also, for generic triangles (e.g. with fermions
in the loop or with vertices inclusing derivatives) one needs to redo the analysis of \cite{tHooft:1978jhc}
for each specific problem. 

In this work we present a simple approach which allows one to obtain numerically the triangle diagram at any $p_i^2$,
including the timelike region above the cuts, by evaluating the corresponding Feynman-parameter integral.  

In the Euclidean region, $p_i^2<0$, the integration over the Feynman parameters runs over segments
of the real axis. We demonstrate (by extensive comparisons with the known rigorius results from dispersion representations
and an explicit analysis of migration of singularities)
that the same feature holds also in the physical Minkowski region of the external momenta above the cuts:
The integration over the Feynman parameters still runs along the same segments of the real axis. No deformation
of the integration paths is forced by possible migrations of poles from the unphysical sheet
to the physical one. Ultimately, the triangle diagram in the physical Minkowski region above the cuts is given
by the same Feynman-parameter integral for $F(p_1^2,p_2^2,p_3^2|m_1,m_2,m_3)$
through the replacements $p_i^2 \to p_i^2 + i0$.

Our observation opens the possibility to obtain the generic triangle diagram (e.g.
with fermions in the loop and with vertices containing derivatives) in the Minkowski region above the cuts
numerically by a simple evaluation of the Feynman-parameter integral over the segments of the real axis.

\section{Triangle diagram and its Feynman-parameter integral representation}

\begin{figure}[t!]
\begin{center}
\includegraphics[width=4.5cm]{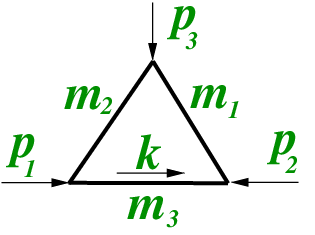} 
\caption{\label{fig:0} 
The Feynman diagram for $F(p_1^2,p_2^2,p_3^2|m_1,m_2,m_3)$.}
\end{center}
\end{figure}

We shall illustrate our approach using an example of one-loop triangle diagram with spinless particles
in the loop (Fig.~\ref{fig:0}). 
It has the following analytic form  
\begin{equation}
\label{f}
F(p_1^2,p_2^2,p_3^2|m_1,m_2,m_3)=\frac{1}{(2\pi)^4 i}\int  
\frac{dk}{(m_3^2-k^2-i0)(m_2^2-(p_1-k)^2-i0)(m_1^2-(p_2+k)^2-i0)} 
\end{equation}
with $p_1+p_2+p_3 = 0$.
The integral (\ref{f}) is a contour integral and requires handling with care.
We now discuss our procedure step-by-step. 

\noindent
(a) The first step is quite standard. One considers all $p_i^2<0$ in the Euclidean,
  introduces Feynman parameters to join three propagators
  into the third power of the quadratic polynomial in the denominator, performs the Wick rotation
  of the integration contour and, finally, takes the integral over $dk_E$. As the result of this step,
  one obtains the following Feynman-parameter representation valid at $p_i^2<0$ (i=1,2,3) 
\begin{eqnarray}
\label{fp}
&&F(p_1^2,p_2^2,p_3^2|m_1,m_2,m_3)=\frac{1}{16\pi^2}
\int\limits_0^1 du_1\int\limits_0^{1-u_1} du_2\\
&&\qquad\qquad\times\frac{1}{m_1^2u_1+m_2^2 u_2+m_3^2(1-u_1-u_2)-p_1^2 u_2(1-u_1-u_2)-p_2^2 u_1(1-u_1-u_2)- p_3^2 u_1u_2}.\nonumber 
\end{eqnarray}
Recall that the $-i0$ terms in the Feynman propagators in (\ref{f}) were used in order to properly perform the Wick rotation.
No $i0$ term in the denominator is present anymore; the function $F(p_1^2,p_2^2,p_3^2|m_1,m_2,m_3)$ is
a real function in the Euclidean region of all external momenta. The integral (\ref{fp}) is a usual
definite integral. 

One can perform the integration over the Feynman parameters and obtain the explicit analytic expression
for the triangle in the Euclidean region in terms of the dilogarithmic function \cite{tHooft:1978jhc}. One can then
perform the analytic continuation to the Minkowski region using this explicit expression. This is however not an easy task
because one needs to stay at proper branches of a multi-valued dilogarithmic function. For a scalar trianlge
\cite{tHooft:1978jhc} gives an explicit solution. For a generic non-scalar triangle diagram the analysis should be
performed on a case-by-case basis.

Our goal here is different: Without performing the integration over the Feynman parameters in the Euclidean region,
we are going to see what happens with the Feynman-parameter integral (\ref{fp}) if the variables $p_i^2$ are
moved to the Minkowski region. 

\noindent
(b) The integral (\ref{fp}) cannot be used as it is in the timelike region $p_i^2>0$ because of the appearance of
poles in the integration region. This feature leads to the cuts along the real axes of the variables $p_i^2$.
In order to evaluate $F(p_1^2,p_2^2,p_3^2|m_1,m_2,m_3)$ in the physical Minkowski region above the cut, 
let us try to make the following replacement in the Feynman-parameter integral (\ref{fp})  
\begin{eqnarray}
\label{pre}
 p_i^2\to p_i^2+i\epsilon \quad \mbox{(or, equivalently, } m_i^2\to m_i^2-i\epsilon).
\end{eqnarray}
Knowing the analytic properties of the form factor, one may anticipate that this simple replacement gives the
form factor on the upper boundary of the cut. But we emphasize that at this stage, before a detailed analysis
has been performed, this is an {\it extra step} which modifies the Feynman-parameter integral. 

As soon as $i\epsilon$ appears in the denominator, the integral over the Feynman parameters should be
treated as a contour integral in the complex plane of Feynman parameters $u_1$ and $u_2$.
The starting point is that the parameter $u_1$ runs along the segment of the real axis $[0,1]$,
whereas the parameter $u_2$ runs along the segment 
$[0,1-u_1]$. Since the Feynman-parameter integral is now a contour integral,
one has to study the migration of poles of the
denominator depending on the values of $p_i^2$, $m_i^2$, $u_1$ and $u_2$. 

The central result of our analysis is the following one.
{\it No deformation of the $u_1$ and $u_2$ integration paths along the real axis 
  is forced by the migration of the poles}.\footnote{Unless there is pinching of
singularities, as it occurs e.g.\,at the anomalous thresholds
(see Appendix \ref{AppendixA}),
one has the freedom to deform the contour in order
to improve the numerical integration as it was proposed in \cite{Anastasiou:2005cb,Nagy:2006xy,Anastasiou:2007qb}.
We underline however that no deformation is forced by the migration of singularities.}
This makes an essential difference with the appearance of the anomalous cut 
in a single dispersion representation, forced by the migration of the singularity from
the unphysical sheet onto the physical sheet through the normal cut. 
Thus, one can safely keep the $u_1$-integration running along the segment of the real axis $[0,1]$ and the
$u_2$-integration running along the segment of the real axis $[0,1-u_1]$. This allows a straightforward
numerical evaluation of the Feynman-parameter integral for any $p_i^2$, both spacelike and timelike.

\noindent
(c)
As is well known, the form factor has cuts and thresholds, both normal and anomalous. We shall demonstrate
  through extensive comparisons with the 
known results from dispersion representations, that the Feynman-parameter integral (\ref{fp}) complemented
by the prescription (\ref{pre}) indeed provides the analytic function with properly located singularities
(the anomalous threshold with a logarithmic singularity and the normal threshold with a square-root singularity)
on the upper boundary of the cut.  

Our findings allow one to obtain numerical predictions for the form factors at arbitrary
values of the external momenta and masses in the loop by evaluating the Feynman-parameter integral
over a simple real integration region.

In the numerical integration one keeps $\epsilon \ne 0$ and then needs to study the limit $\epsilon \to 0$. 
For $\epsilon \ne 0$, one obtains a smooth finite function on the upper boundary of the cut.  
The singularities of the form factor at the anomalous and the normal thresholds
are recovered as soon as one takes $\epsilon \to 0$, 
because of pinching of poles in $u_1$ and $u_2$ complex planes.
Let us emphasize once more that these pinches do not force any deformation of the integration contours
which are still the segments of the real axis in $u_1$ and $u_2$
(see the proof in Appendix \ref{AppendixA}). 

Beyond the thresholds $\epsilon$ may be taken as small as dictated by the necessary accuracy of the form factor.
Near the singularities the accuracy of the numerical predictions is clearly limited. But, this is a standard
situation for regulating a singularity and therefore no strong practical limitations arise from this feature.

We have checked our numerical results by using also the $\lambda$-method for numerical integration presented in \cite{Anastasiou:2007qb}, obtaining an excellent agreement in all the cases considered.


\section{\label{sect:ii}Single spectral representation}
Let us start with the case of the elastic form factor: particles of the same mass propagate in the loop,
$m_1=m_2=m_3=m$, $p_1^2=p_2^2=M^2$ and $p_3^2\equiv q^2$. So, we denote 
$F_{\rm el}(q^2,M^2)\equiv F(M^2,M^2,q^2|m,m,m)$.  
We start with $M^2<0$ where a single dispersion representation in $q^2$ may be written as \cite{lms,dm_particles}
\begin{eqnarray}
\label{single}
F_{\rm el}(q^2,M^2)=\frac{1}{\pi}\int \frac{dt}{t-q^2-i0}\sigma(t,M^2).    
\end{eqnarray} 
The spectral density in the region $M^2<0$ is given by Cutkosky rules:  
\begin{eqnarray}
\label{normMle0}
\sigma_{\rm norm}(t,M^2)=
\frac1{16\pi\sqrt{t(t-4M^2)}}
\log\left(
\frac{t-2M^2+\sqrt{t(t-4M^2)}\sqrt{1-4m^2/t}}{t-2M^2-\sqrt{t(t-4M^2)}\sqrt{1-4m^2/t}}
\right), \qquad M^2<0.
\end{eqnarray}
It turns out that this expression is also valid in a broader domain, namely, $M^2<m^2$.  
Performing the analytic continuation in variable $M^2$ from the domain $M^2<m^2$, we obtain the spectral density 
for $2m^2<M^2<4m^2$: 
\begin{eqnarray}
\label{normM}
\sigma_{\rm norm}(t,M^2)=
\left\{\begin{array}{ll} 
\frac1{16\pi\sqrt{t(t-4M^2)}}
\log\left(
\frac{t-2M^2+\sqrt{t(t-4M^2)}\sqrt{1-4m^2/t}}{t-2M^2-\sqrt{t(t-4M^2)}\sqrt{1-4m^2/t}}
\right), 
& \quad
4M^2 \le t,
\\ 
\frac1{8\pi
\sqrt{-t(t-4M^2)
}}\arctan
\left(
\frac{\sqrt{t(4M^2-t)}\sqrt{1-4m^2/t}}{t-2M^2}
\right), 
& 
\quad  
2M^2 \le t\le 4M^2,
\\ 
\frac1{8\pi
\sqrt{-t(t-4M^2)
}}
\left[\pi+\arctan\left(
\frac{\sqrt{t(4M^2-t)}\sqrt{1-4m^2/t}}{t-2M^2}
\right)\right], 
& 
\quad  4m^2\le t\le 2M^2.
\end{array}\right. 
\end{eqnarray}
A peculiar feature of the spectral representation (\ref{single}) is the appearance of the anomalous cut with 
the anomalous threshold at $t_0=\frac{M^2}{m^2}(4m^2-M^2)$ for $M^2>2m^2$. 
The anomalous cut is related to a migration of the singularity at $t=t_0$ of the logarithmic 
function in (\ref{normMle0}). For $M^2\le 2m^2$, $t_0$ is located on the unphysical sheet and does not influence the
dispersion integral. However, for $M^2>2m^2$, $t_0$ goes onto the physical sheet around the normal threshold and requires
the deformation of the integration contour in the dispersion integral. This leads to the appearance of the anomalous cut
from $t_0$ to $4m^2$. The discontinuity of the form factor $F(q^2,M^2)$ on the anomalous cut is related 
to the discontinuity of the function $\sigma_{\rm norm}(t,M^2)$ and reads \cite{dm_particles}
\begin{eqnarray}
\label{anomM}
\sigma_{\rm anom}(t,M^2)=
\frac{1}{8\sqrt{-t(t-4M^2)}},\qquad t_0\le t\le 4m^2, \qquad t_0=\frac{M^2}{m^2}(4m^2-M^2).
\end{eqnarray} 
Therefore, the full spectral density has the form 
\begin{eqnarray}
\sigma(t,M^2)=
\theta(M^2-2m^2)
\theta(t_0\le t\le 4m^2)\sigma_{\rm anom}(t,M^2)
+
\theta(4m^2\le t)\sigma_{\rm norm}(t,M^2). 
\end{eqnarray}
Clearly, the spectral density given by Eqs.~(\ref{normM}) and (\ref{anomM}) is a continuous function for 
$t>t_0$. Moreover, the anomalous spectral density does not vanish at the anomalous threshold $t_0$, thus leading to
a logarithmic singularity of the form factor at $q^2=t_0$ of the form $\log(q^2-t_0)$. 

The spectral representation for the form factor contains the anomalous and the normal contributions:
\begin{eqnarray}
\label{singledisp}
F_{\rm el}(q^2,M^2)=
\theta(M^2-2m^2)
\int\limits_{t_0}^{4m^2}
\frac{dt}{\pi(t-q^2-i0)}\sigma_{\rm anom}(t,M^2)
+
\int\limits_{4m^2}^{\infty}
\frac{dt}{\pi(t-q^2-i0)}\sigma_{\rm norm}(t,M^2). 
\end{eqnarray}


\begin{figure}[!t]
  \begin{center}
    \begin{tabular}{cc}
      \includegraphics[width=8cm]{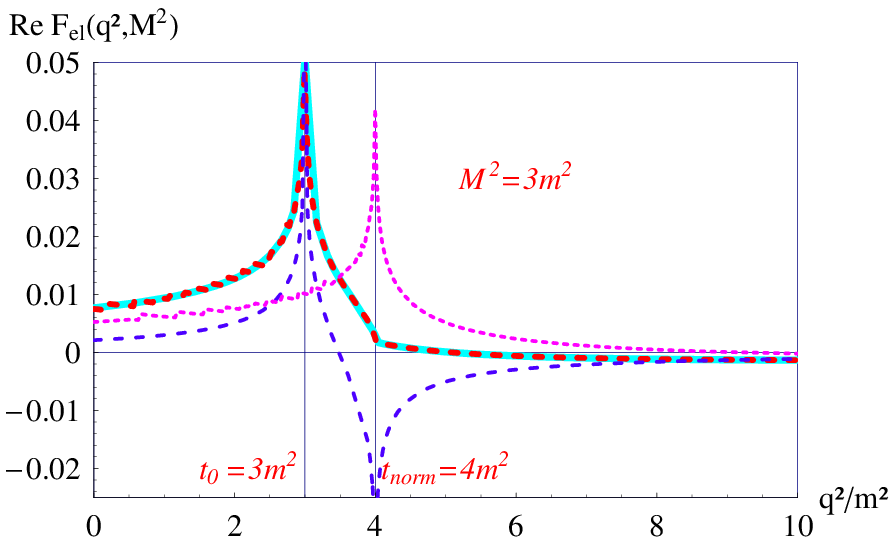} &
      \includegraphics[width=8cm]{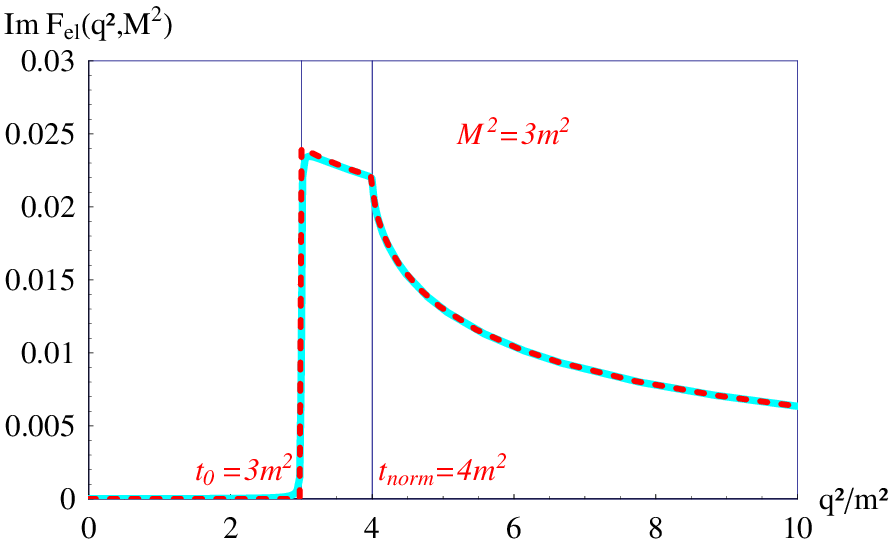}\\
      (a)   &    (b)
      \end{tabular}
   \caption{\label{fig:1}The single dispersion representation vs the Feynman parameter representation
     for the real (a) and imaginary (b) parts of $F_{\rm el}(q^2,M^2)$; $M^2=3m^2$, m=1 GeV. Two thresholds are
visible in the real and in the imaginary parts of the form factor $F_{\rm el}(q^2,M^2)$: the anomalous threshold
is at $t_0=3m^2$ and the normal unitary threshold is at $t_{\rm norm}=4m^2$.
Solid light-blue line: the Feynman-parameter representation (\ref{fp}) complemented
by the prescription (\ref{pre}) with $\epsilon/m^2=0.001$.
Dashed blue line: the anomalous contribution to the dispersion representation (\ref{singledisp}). 
Dashed magenta line: the normal contribution to the dispersion representation (\ref{singledisp}). 
Dotted red line: the sum of the anomalous and the normal contributions in (\ref{singledisp}).}    
\end{center}
\end{figure}
Figure \ref{fig:1} shows the calculated $F_{\rm el}(q^2,M^2)$ for the case of the anomalous kinematics $2m^2<M^2<4m^2$,
which leads to the anomalous threshold at $t_0=\frac{M^2}{m^2}(4m^2-M^2) > 0$. We present the results for the anomalous
and the normal contributions to $F_{\rm el}(q^2,M^2)$, for the case $m=1$ GeV, $M^2=3m^2$, for which the anomalous threshold
$t_0=3m^2$ is below the normal threshold at $t_{\rm norm}=4m^2$. For comparison we present the results of evaluating
the Feynman-parameter integral (\ref{fp}). In the latter we add a small imaginary part to the masses:  
$m^2\to m^2-i\epsilon$. 
The Feynman-parameter integral excellently reproduces the result of the analytic continuation,
including both the normal and the anomalous parts. 

\begin{figure}[!t]
  \begin{center}
    \begin{tabular}{cc}
      \includegraphics[width=8cm]{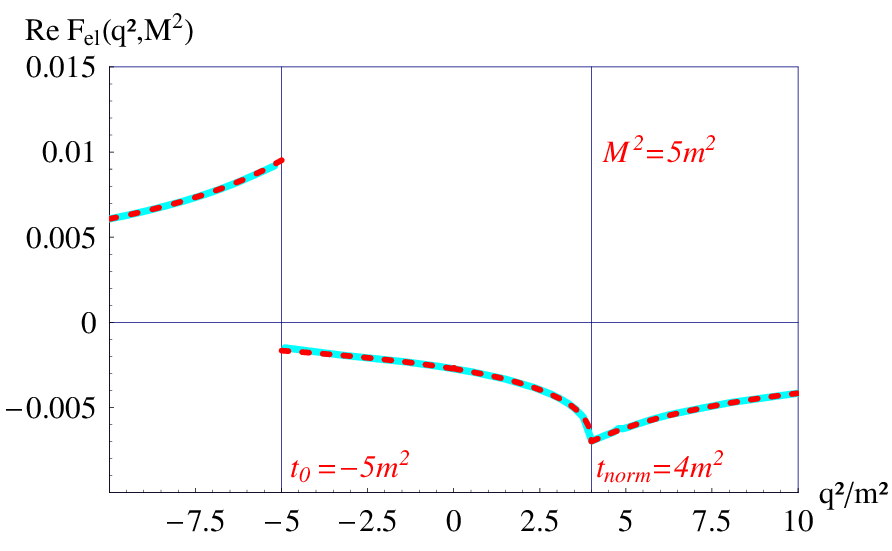} &
      \includegraphics[width=8cm]{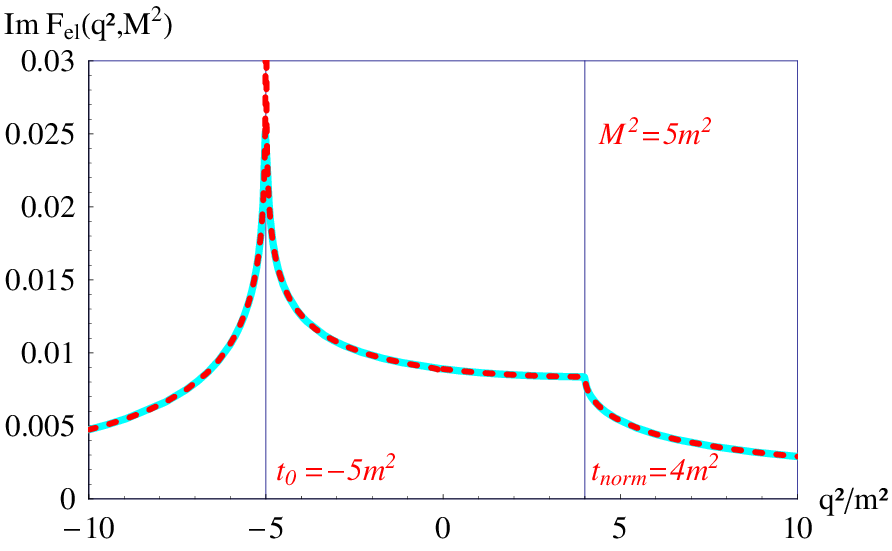}\\
      (a)   &    (b)
      \end{tabular}
   \caption{\label{fig:1b}The single dispersion representation vs the Feynman parameter representation
     for the real (a) and imaginary (b) parts of $F_{\rm el}(q^2,M^2)$ for $M^2$ above the two-particle threshold,
     $M^2=5m^2$, m=1 GeV. Two thresholds are clearly visible in the real and in the imaginary parts of the
     form factor $F_{\rm el}(q^2,M^2)$: The anomalous threshold is at $t_0=-5m^2$ and the normal unitary
     threshold is at $t_{\rm norm}=4m^2$.
     Solid light-blue line: the Feynman-parameter representation (\ref{fp}) complemented
     by the prescription (\ref{pre}) with $\epsilon/m^2=0.001$.
Dotted red line: the single dispersion representation (\ref{singledisp}) with (\ref{sigmaanom2}).}
   \end{center}
\end{figure}

Our consideration above was limited to the case $M^2<4m^2$. 
Let us now consider the case when $M^2$ is above its two particle threshold, $M^2>4m^2$,  
and $t_0$ becomes negative. For the application of the dispersion representation
(\ref{singledisp}) we have to determine $\sigma_{\rm anom}(t)$ for $t<0$. A direct analytic
continuation gives
\begin{eqnarray}
\label{sigmaanom2}
\sigma_{\rm anom}(t)=\frac{i}{\sqrt{-t(4M^2-t)}},\qquad t<0. 
\end{eqnarray}
We can then obtain $F_{\rm el}(q^2,M^2)$ for all values of $q^2$ using both the single spectral representation
(\ref{singledisp}) and the Feynman-parameter representation (\ref{fp}). Figure~\ref{fig:1b} shows an excellent
agreement between the results of both representations. 

Before closing this Section, we make the following remark. The spectral density (\ref{sigmaanom2})
is imaginary and this is rather unusual. We test this feature as follows. The double spectral
representation, discussed in the next Section, provides a direct and unambiguous representation
of $F_{\rm el}(q^2,M^2)$ for $q^2<0$ and any $M^2$, both below and above the threshold $4m^2$.
We shall see below in Fig.~\ref{fig:3} that the results of the
single spectral representation, of the double spectral representation and of the Feynman-parameter integral provide
precisely the same result for the form factor $F_{\rm el}(q^2,M^2)$ for $q^2<0$. This confirms that the
single spectral density in (\ref{anomM}) and (\ref{sigmaanom2}) is defined properly.

\section{\label{sect:iv}Double spectral representation}

Now we discuss the triangle diagram with particles of different masses in the loop, $m_2=\mu$, $m_1=m_3=m<\mu$, 
$p_1^2\ne p_2^2$, $p_3^2\equiv q^2$, and consider the decay kinematics $0<q^2<(\mu-m)^2$ \cite{braun,melikhov}.
We denote the corresponding weak transition form factor by $F_{\rm weak}(q^2,p_1^2,p_2^2)\equiv F(p_1^2,p_2^2,q^2|m,\mu,m)$. 

\subsection{Transition form factor at $q^2<0$}
We start with the Euclidean region $q^2<0$, where the double dispersion representation has the form \cite{melikhov} 
\begin{equation}
\label{fftrans2}
F_{\rm weak}(q^2,p_1^2,p_2^2)=\int\limits^\infty_{4m^2}\frac{ds_2}{\pi(s_2-p_2^2)}
\int\limits^{s_1^+(s_2,q^2)}_{s_1^-(s_2,q^2)}\frac{ds_1}{\pi(s_1-p_1^2)}
\frac{1}{16\lambda^{1/2}(s_1,s_2,q^2)}, 
\end{equation}
with 
\begin{eqnarray}
\label{s1pm}
s_1^\pm(s_2,q^2)&=&
\frac{s_2(m^2+\mu^2-q^2)+2 m^2 q^2}{2m^2}
\pm\frac{\lambda^{1/2}(s_2,m^2,m^2)\lambda^{1/2}(q^2,\mu^2,m^2)}{2m^2}, \qquad \lambda(a,b,c)\equiv (a-b-c)^2-4bc.\nonumber\\
\end{eqnarray}
A new feature compared with the case of equal masses in the loop  
is the appearance of the region $0<q^2<(\mu-m)^2$, which was absent in the equal-mass case.  
This region corresponds to the decay of a particle of mass $\mu$ to
a particle of mass $m$ with the emission of a particle of mass $\sqrt{q^2}$. 

\subsection{Transition form factors at $q^2>0$}
The form factor in the region $0<q^2<(\mu-m)^2$ may be obtained by analytic 
continuation of the expression (\ref{fftrans2}). 
Let us consider the structure of the singularities of the integrand in Eq.~(\ref{fftrans2}) 
in the complex $s_1$-plane for a fixed 
real value of $s_2$ in the interval $s_2>4m^2$. 

The integrand has singularities (branch points) 
related to the zeros of the function $\lambda(s_1,s_2,q^2)$ at $s_1^L=(\sqrt{s_2}-\sqrt{q^2})^2$ and 
$s_1^R=(\sqrt{s_2}+\sqrt{q^2})^2$. As $q^2\le 0$, these singularities lie on the unphysical sheet. 
However, as $q^2$ becomes positive, the point $s_1^R$ may move onto the physical sheet through the cut 
from $s_1^-$ to $s_1^+$. This happens for values of the variable $s_2>s_2^0$, with 
$s_2^0$ obtained as the solution to the equation $s_1^R(s_2,q^2)=s_1^-(s_2,q^2)$. Explicitly, 
one finds  
\begin{eqnarray}
\sqrt{s_2^0}=\frac{\mu^2-m^2-q^2}{\sqrt{q^2}}.
\end{eqnarray}
As $q^2>0$, for $s_2>s_2^0(q^2)$ the
integration contour in the complex $s_1$-plane should be deformed such that it embraces the 
points $s_1^R$ and
$s_1^+$. Respectively, the $s_1$-integration contour contains two segments: the normal part from 
$s_1^-$ to $s_1^+$, and the anomalous part from $s_1^R$ to $s_1^-$. The double spectral density 
for the anomalous piece is just the discontinuity of the function 
$1/\sqrt{\lambda(s_1,s_2,q^2)}$. It can be easily calculated as follows: 
Recall the relation $\sqrt{\lambda(s_1,s_2,q^2)}=\sqrt{s_1-s_1^L}\sqrt{s_1-s_1^R}$. The branch point 
$s_1^L$ lies on the unphysical sheet, therefore the function $\sqrt{s_1-s_1^L}$ is continuous 
on the anomalous cut located on the physical sheet. Thus, we have to calculate the 
discontinuity of the
function $1/\sqrt{s_1-s_1^R}$ which is twice the function itself. As the result, the discontinuity
of the function $1/\sqrt{\lambda(s_1,s_2,q^2)}$ on the anomalous cut is $2/\sqrt{\lambda(s_1,s_2,q^2)}$.
Finally, the full double spectral density including the normal and the anomalous pieces 
takes the form (see e.g. \cite{dm_particles}) 
\begin{eqnarray}
\label{4deltas}
\Delta(q^2,s_1,s_2|\mu,m,m)&=&
\frac{\theta(s_2-4m^2)\theta(s_1^-<s_1<s_1^+)}{16\lambda^{1/2}(s_1,s_2,q^2)}
+\frac{2\theta(q^2)\theta(s_2-s_2^0)\theta(s_1^R<s_1<s_1^-)}{16\lambda^{1/2}(s_1,s_2,q^2)}.
\end{eqnarray}
The first term in (\ref{4deltas}) relates to the Landau-type contribution
emerging when all
intermediate particles go on mass shell, while the second term describes the
anomalous contribution.

The final representation for the form factors at $0<q^2<(\mu-m)^2$
takes the form  
\begin{eqnarray}
\label{final1}
F_{\rm weak}(q^2,p_1^2,p_2^2)&=&
\int\limits_{4m^2}^\infty\frac{ds_2}{\pi(s_2-p_2^2-i0)}
\int\limits_{s_1^-(s_2,q^2)}^{s_1^+(s_2,q^2)}
\frac{ds_1}{\pi(s_1-p_1^2)}
\frac{1}{16\lambda^{1/2}(s_1,s_2,q^2)}
\\
&&+
2\theta\left(0<q^2<(\mu-m)^2\right)\int\limits_{s_2^0(q^2)}^\infty\frac{ds_2 }{\pi(s_2-p_2^2-i0)}
\int\limits_{s_1^R(s_2,q^2)}^{s_1^-(s_2,q^2)}
\frac{ds_1}{\pi(s_1-p_1^2)}\frac{1}{16\lambda^{1/2}(s_1,s_2,q^2)}.\nonumber
\end{eqnarray}
We point out that the representation (\ref{final1}) is particularly suitable for application to processes 
where $p_1^2$ and $p_2^2$ are above two-particle thresholds: the spectral densities do not depend on $p_1^2$
and $p_2^2$, so for values of $p_1^2$ and $p_2^2$ above the thresholds one just 
has to take into account the appearance of the absorptive parts in the $s_1$ and $s_2$ integrals.

\begin{figure}[!ht]
\begin{center}
  \includegraphics[width=8cm]{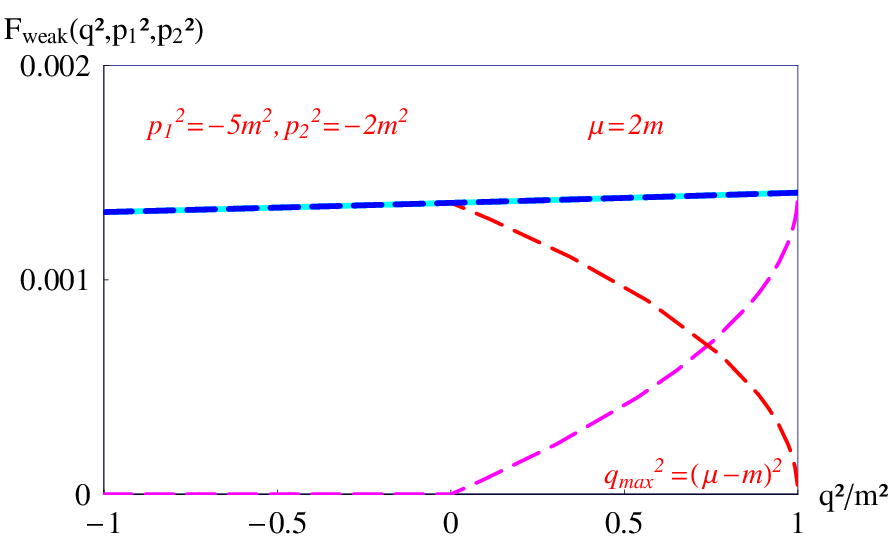}
  \caption{\label{fig:2}The double dispersion representation vs the Feynman-parameter representation
     for $F_{\rm weak}(q^2,p_1^2,p_2^2)$ with $\mu=2m$, $p_1^2=-5m^2$, $p_2^2=-2m^2$.
Solid light-blue line: the Feynman-parameter representation (\ref{fp}) with (\ref{pre}).
Dashed magenta line: the anomalous contribution to the double dispersion representation (\ref{final1}).
Dashed red line: the normal contribution to the double dispersion representation (\ref{final1}). 
Dashed blue line: the sum of the anomalous and the normal contributions in (\ref{final1}).}    
\end{center}
\end{figure}

Figure \ref{fig:2} compares the Feynman-parameter representation for the triangle diagram and the
double-dispersion integral (\ref{final1}) in the region $q^2<(\mu-m)^2$ and for $p_1^2=-5m^2$ and $p_2^2=-2m^2$. 
In this case the form factor is real, but its double dispersion representation
contains the anomalous and the normal parts. The exact result of the double-dispersion
representation at $p_1^2$ and $p_2^2$ on the upper boundary of the cuts is reproduced by
the Feynman-parameter representation by adding in the denominator of (\ref{fp}) the imaginary parts
\begin{eqnarray}
m^2\to m^2-i\epsilon, \quad \mu^2\to \mu^2-i\epsilon.  
\end{eqnarray}

\section{Feynman-parameter representation above thresholds}
We now give examples showing that the exact results of the analytic continuation
of the dispersion representations for the form factor to the Minkowski region
are reproduced by the Feynman-parameter integral (\ref{fp}) with
the condition (\ref{pre}). In doing that we consider Feynman-parameter representation for the form factor 
in a broad region of external momenta where direct application of the dispersion representations
may be difficult.
We emphasize that the Feynman-parameter
representation in all cases under discussion is evaluated for momenta squared {\it above the real axis}
\begin{eqnarray}
F(q^2+i0,p_1^2+i0,p_2^2+i0). 
\end{eqnarray}
\begin{figure}[!h]
  \begin{center}
    \begin{tabular}{cc}
      \includegraphics[width=8cm]{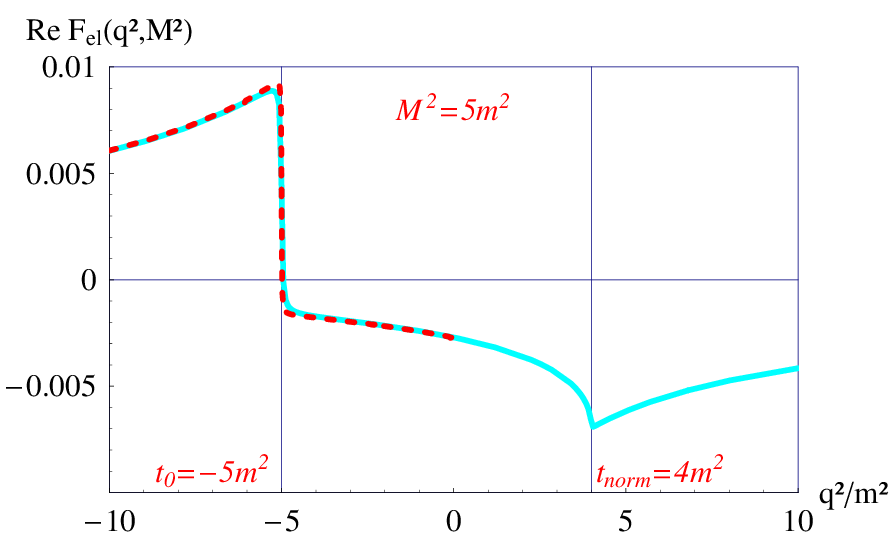} & 
      \includegraphics[width=8cm]{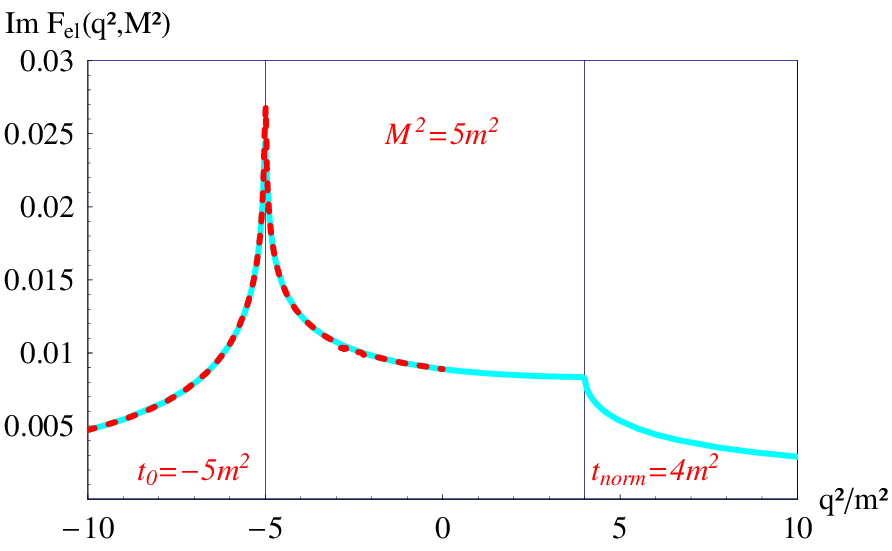}\\
      (a)  &  (b)
      \end{tabular}
    \caption{\label{fig:3}
      The form factor $F_{\rm el}(q^2,M^2)$ vs $q^2$ for $M^2=5m^2$, $\epsilon/m^2=0.001$.
     Solid light-blue-line - the Feynman-parameter integral (\ref{fp});
     Dashed red line - the double dispersion representation (\ref{final1}).  
     The anomalous threshold $t_0=-5m^2$ and the normal threshold $t_{norm}=4m^2$ 
     are clearly visible in both the real (a) and in the imaginary (b) parts of the form factor.}
    \begin{tabular}{cc}
      \includegraphics[width=8cm]{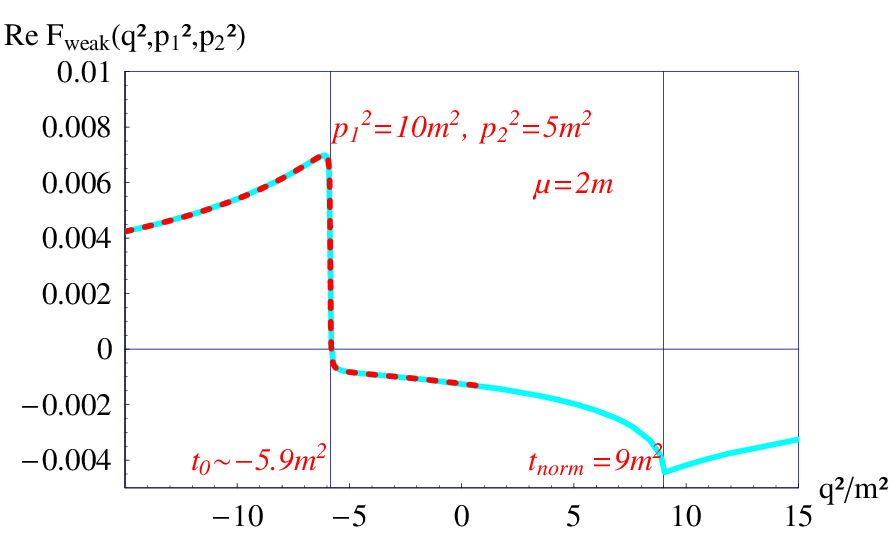} & 
      \includegraphics[width=8cm]{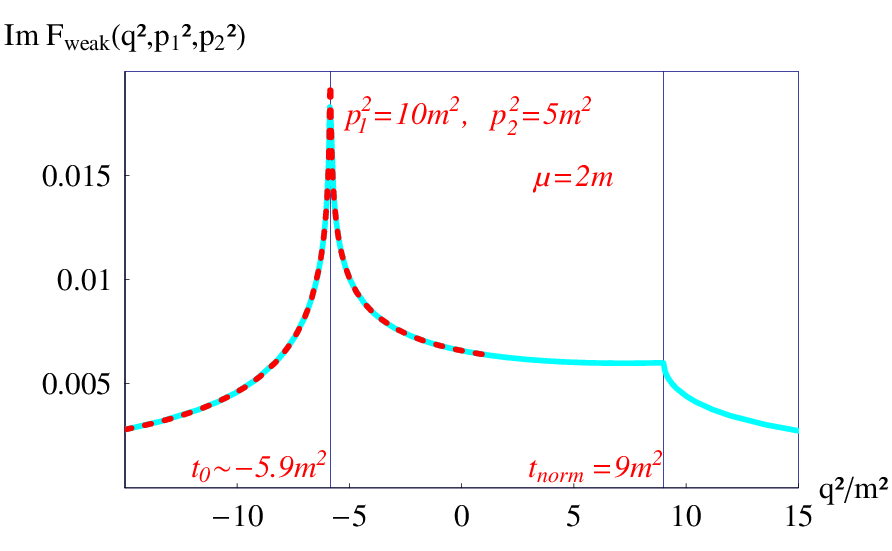}
      \end{tabular}
   \caption{\label{fig:4}The form factor $F_{\rm weak}(q^2,p_1^2,p_2^2)$ vs $q^2$ for $\mu=2m$,
     $p_1^2=10m^2$, $p_2^2=5m^2$, $\epsilon/m^2=0.001$.
     Solid light-blue-line - the Feynman-parameter integral (\ref{fp});
     Dashed red line - the double dispersion representation (\ref{final1}).  
     The anomalous threshold $t_0=\simeq -5.9 m^2$ and the normal threshold $t_{norm}=9m^2$ 
     are clearly visible in both the real (a) and in the imaginary (b) parts of the form factor.}
   \begin{tabular}{cc}
      \includegraphics[width=8cm]{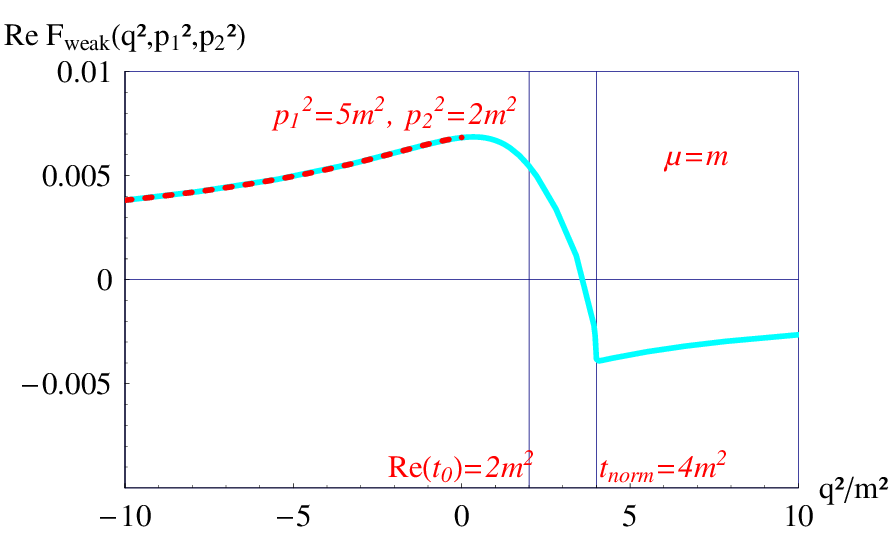} & 
      \includegraphics[width=8cm]{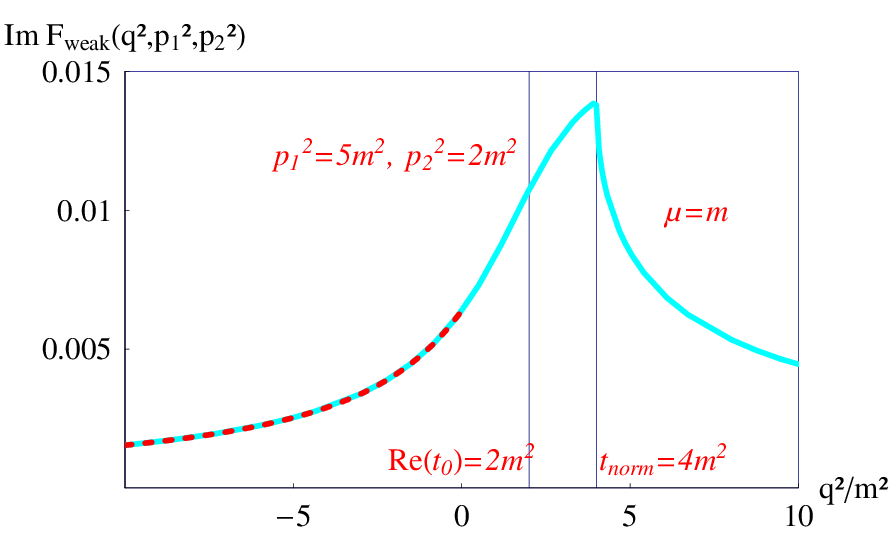}
      \end{tabular}
   \caption{\label{fig:5}
     The form factor $F_{\rm weak}(q^2,p_1^2,p_2^2)$ vs $q^2$ for $\mu=m$,
     $p_1^2=5m^2$, $p_2^2=2m^2$, $\epsilon/m^2=0.001$.
     Solid light-blue-line - the Feynman-parameter integral (\ref{fp});
     Dashed red line - the double dispersion representation (\ref{final1}).
     The normal threshold $t_{norm}=4m^2$ is clearly visible in both the real (a) and in the imaginary (b)
     parts of the form factor. The anomalous threshold lies far from the real axis at $t_0=2m^2(1-i\sqrt5)$
     and is not visible in the form factor for real values of $q^2$.} 
\end{center}
\end{figure}

\subsubsection{Elastic form factor}
Figure \ref{fig:3} shows the Feynman-parameter representation for the
elastic form factor $F_{\rm el}(q^2,M^2)$ for 
$m=1$ GeV and $p_1^2=p_2^2=M^2=5m^2$ above the two-particle threshold. 
The elastic form factor in this case contains real and imaginary parts. 
One can clearly see the normal threshold at $t_{\rm norm}=4m^2$ and the anomalous threshold
at
\begin{eqnarray}
t_0=\frac{M^2}{m^2}(4m^2-M^2)=-5m^2
\end{eqnarray}
both in the imaginary and in the real parts of the form factor
$F_{\rm el}(q^2,M^2)$. As anticipated in Section 3, the results shown in Fig.~\ref{fig:3} 
coincide also with those obtained using the single dispersion representation (\ref{singledisp})
complemented with (\ref{sigmaanom2}), see Fig.~\ref{fig:1b}.

\subsubsection{Weak-transition form factor}
Figure \ref{fig:4} illustrates the behaviour of the Feynman-parameter representation
for the weak-transition form factor $F_{\rm weak}(q^2,p_1^2,p_2^2)$ for $\mu=2m$, $m=1$ GeV, 
$p_1^2=10m^2$, $p_1^2=5m^2$. Both $p_1^2$ and $p_2^2$ lie above their
two-particle threshold at $p_1^2=(\mu+m)^2$ and $p_2^2=4m^2$.
The anomalous threshold in $q^2$ emerges for $p_1^2$ and $p_2^2$ satisfying 
\begin{eqnarray}
p_1^2+\frac{\mu}{m}p_2^2>(m+\mu)^2.
\end{eqnarray}
The location of the anomalous threshold is given by the Landau equations and has the form 
\begin{eqnarray}
\label{t0}
t_0=\frac{2m^2p_1^2+(m^2+\mu^2)p_2^2-p_1^2p_2^2}{2m^2}-\frac{\lambda^{1/2}(p_2^2,m^2,m^2)\lambda^{1/2}(p_1^2,\mu^2,m^2)}{2m^2}.
\end{eqnarray}
For the considered masses and external variables, one finds $t_0\simeq -5.9 m^2$.
Both the normal and the anomalous thresholds are clearly seen in the imaginary and the real parts of
the weak-transition form factor.

Figure \ref{fig:5} shows the form factor $F(q^2,p_1^2,p_2^2)$ for the case $\mu=m=1$ GeV,
$p_1^2=5m^2$ above the two-particle threshold and $p_2^2=2m^2$ below the two-particle threshold.
In this case the anomalous threshold (\ref{t0}) lies in the complex plane at $t_0=2m^2(1-i\sqrt5)$
and is not visible in the form factor. 

\subsubsection{Double dispersion representation in the Euclidean region}
We would like to emphasize that in the Euclidean region ($q^2<0$ for the elastic case and
$q^2<(\mu-m)^2$ in the weak-transition case), the double dispersion representation
(\ref{final1}) of the previous Section can be readily applied for any values of $p_1^2$ and $p_2^2$,
including the regions above the two-particle thresholds. As shown in Figs.~\ref{fig:3}--\ref{fig:5}, 
the results obtained from double dispersion representations
excellently agree with the results of the Feynman-parameter integrals
at $q^2<0$ in the elastic case and at $q^2<(\mu-m)^2$ in the weak-transition case.

These tests give us confidence that the Feynman-parameter integral (\ref{fp}) with the additions of the
imaginary parts (\ref{pre}) provides a proper representation of the analytic function
$F(p_1^2,p_2^2,p_3^2|m_1,m_2,m_3)$ in the Minkowski region.


\section{Summary and Conclusions} 
We presented a detailed comparison of Feynman-parameter representation for the triangle diagram with the 
dispersion representations placing emphasis on the kinematical regions of Minkowski external momenta, 
aawasawsaawhere the dispersion representations contain anomalous cuts and thresholds.

1. For the case of the elastic form factor $F_{\rm el}(q^2,M^2)$ we have demonstrated (by numerical calculations)
that the results of the single dispersion representation, of the double dispersion representation and of the Feynman-parameter
representation give precisely the same results at $q^2<0$ for any $M^2$. Moreover, the single dispersion
representation and the Feynman-parameter integral give the same results for all $M^2$ and all $q^2$. 

2. For the case of the weak transition form factor $F_{\rm weak}(q^2,p_1^2,p_2^2)$, we have shown that 
the double dispersion representation at $q^2<0$ coincides with the Feynman-parameter representation
for any $p_1^2$ and $p_2^2$, below and above the corresponding two-particle thresholds. 
One can clearly identify normal and anomalous thresholds in the form factor.
Thus, the Feynman-parameter representation takes properly into account the analytic
properties of the triangle diagram, including normal and anomalous thresholds and cuts.

Our main conclusion may be formulated as follows.

The Feynman-parameter integral with the integration region along the segments of the real axis
gives a proper result for the triangle diagram
\begin{eqnarray}
F(p_1^2,p_2^2,p_3^2|m_1,m_2,m_3) \nonumber 
\end{eqnarray}
not only in Euclidean region of external momenta, but also in Minkowski region $p_i^2>0$ above the cuts.
To obtain the form factor in the latter,
one only needs to introduce extra imaginary parts for the external momenta in the loop:
\begin{eqnarray}
p_i^2\to p_i^2+i\epsilon ~ . ~ \nonumber
\end{eqnarray}
The addition of the infinitesimal positive imaginary parts to the squared momenta corresponds to
the upper boundary of the cuts on the physical sheet. In the numerical calculations we add also negative imaginary parts
to the masses, $m_i^2\to m_i^2-i\epsilon$, which work precisely in the same direction as $p_i^2\to p_i^2+i\epsilon$
and just stabilize the calculations. 

We believe that our result is highly non-trivial. 
As soon as one adds $i\epsilon$ in the denominator of (\ref{fp}),
the integral (\ref{fp}) should be interpreted as a contour integral. 
In the cases of single and double dispersion representations
it is well known that the analytic continuation to the Minkowski region
enforces deformations of the integration contours caused by migration of singularities from
the unphysical sheet onto the physical sheet. {\bf We have seen in our analysis that for the
Feynman-parameter integral no deformation of the integration
region enforced by migrations of singularities is necessary.}
This distinguishes the Feynman-parameter representation from dispersion representations. 

We stress also that performing numerical integration in the
Feynman-parameter representation complemented by the extra replacement $m^2 \to
m^2 - i \epsilon$ selects automatically the proper branch of an analytic function. 
At the same time, the selection of the proper branch of a
multi-valued analytic function becomes a nontrivial problem within fully
analytic approaches. In this respect, the numerical comparison provides
a solid test of our statement that no contour deformation in the
Feynman-parameter space is forced by the migration of singularities.

Therefore, the Feynman-parameter representation provides a convenient tool 
for evaluating triangle diagrams in the Minkowski region of the external momenta $p_i^2$,
including also the cases of fermions in the loop and of complicated vertices in the triangle diagram. 
Moreover, the Feynman-parameter representation may be used as an efficient tool of verifying
the complicated structure of dispersion representations in the timelike region in a
way similar to the analysis carried out in this work.

\appendix
\section{\label{AppendixA} Appearance of the anomalous threshold in the Feynman-parameter integral}

\begin{figure}[h!]
\begin{center}
\includegraphics[width=7cm]{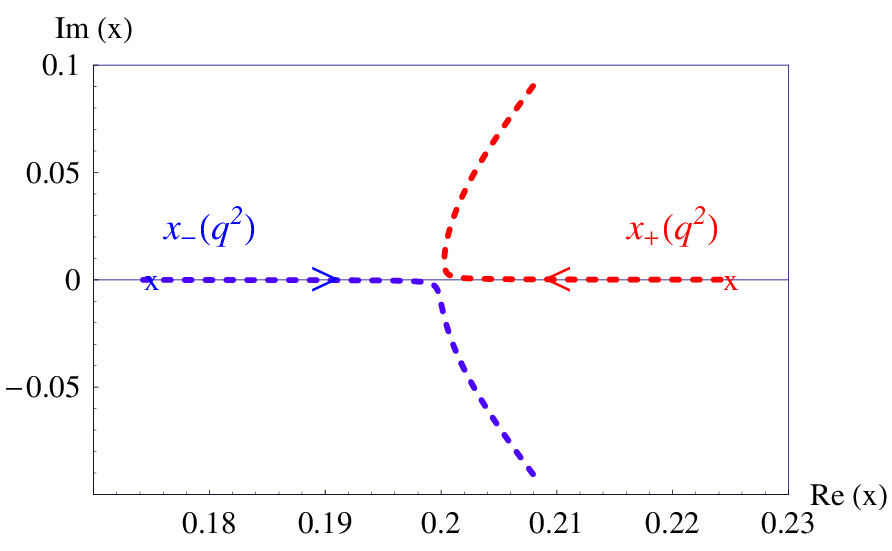} 
\caption{\label{fig:8}
  Trajectories of the singularites $x_\pm(q^2)$ in the complex $x$-plane. Here $M^2=5m^2$, $q_0^2=-5m^2$, and
  the plot is given for $\epsilon=2\cdot 10^{-5}$. The points marked with crosses (x) correspond to $q^2=-6m^2$,
  the arrows indicated the trajectory of $x_\pm$ with increasing $q^2$.
  The pinch occurs at one point $q^2=q_0^2=-5m^2$.}
\end{center}
\end{figure}

To illustrate the appearance of the anomalous singularity
of the triangle diagram in the Feynman-parameter representation,
we perform the integration in (\ref{fp}) over $u_2$. One can easily obtain the general expression for all
different quark masses and external variables, but for the sake of argument we present here the expression for
$p_1^2=p_2^2=M^2$, $m_1=m_2=m_3=m$, and $M^2=5m^2$, corresponding to the case of Fig.~\ref{fig:3} (here $x\equiv u_1$):
\begin{eqnarray}
  F_{\rm el}(q^2,M^2)=\frac{1}{8\pi^2}\int\limits_0^{1}dx(1-x)
\frac{ \arctan\left(N_2/\sqrt{A}\right)- \arctan\left(N_1/\sqrt{A}\right)  }{\sqrt{A}}
\end{eqnarray}
where 
\begin{eqnarray}
N_1&=&-M^2 + 2 M^2 x - q^2 x,\nonumber\\
N_2&=&M^2- q^2 x,\nonumber\\
A&=&4 m^2 M^2 - M^4 - 2 M^2 q^2 x + (4M^2 - q^2)q^2 x^2.
\end{eqnarray}
The function $A$ in the denominator is a quadratic polynomial in $x$ with two roots $x_\pm$:
\begin{eqnarray}
  \label{xpm}
  x_\pm=\frac{M^2q^2\mp 2\sqrt{M^2q^2(M^4+m^2(q^2-4M^2))}}{(4M^2-q^2)q^2}.
\end{eqnarray}
To obtain the form factor at the upper boundary of the cut, we replace $q^2\to q^2+i\epsilon$
(or, equivalently, $m^2\to m^2-i\epsilon$). Taking into account this imaginary addition, we plot
the trajectory of $x_\pm(q^2)$ in the complex $x$-plane in Fig.~\ref{fig:8}.

Two remarks are in order. \\
1. None of the roots cross the real axis: $x_+$ stays above the real axis, whereas $x_-$ stays below the real axis.
This explains why no deformation of the $x$-contour is enforced by the migration of singularities vs $q^2$.\\
2. There is a $q^2$-value at which $x_+$ and $x_-$ pinch. This happes when the square root in (\ref{xpm}) vanishes.
The corresponding value $q_0^2$ is just the anomalous threshold. Using the explicit expression (\ref{xpm}) we find
\begin{eqnarray}
q_0^2=\frac{M^2}{m^2}(4m^2-M^2). 
\end{eqnarray}
At $q^2=q_0^2$, the anomalous logarithmic singularity of the form factor develops.
For nonzero $\epsilon$, one finds
\begin{eqnarray}
F_{\rm el}(q_0^2,M^2,\epsilon)\sim \log(\epsilon) \mbox{ for }\epsilon\to 0. 
\end{eqnarray}
Numerically, the logarithmic type of the singularity may be demonstrated by plotting 
\begin{eqnarray}
 \epsilon \frac{\partial F_{\rm el}(q_0^2,M^2,\epsilon)}{\partial \epsilon} 
\end{eqnarray}
which has a finite value for $\epsilon\to 0$. 

For $q^2\ne q_0^2$, the two roots $x_\pm$ are separated from each other and no singularity in the form factor emerges.

\acknowledgments
The work of D.M. was conducted under the state assignment of Lomonosov Moscow State University.


\end{document}